# A Theoretical Computer Science Perspective on Free Will

Manuel Blum[1] and Lenore Blum[2]

**Abstract**

We consider the paradoxical concept of free will in a deterministic world - from the perspective of **Theoretical Computer Science (TCS)**, a branch of mathematics concerned with understanding the underlying principles of computation and complexity, including the implications and surprising consequences of resource limitations. Although a deterministic entity living in a deterministic world cannot have free will as it is typically understood, it is entirely possible for such an entity to rightly and firmly believe in its own free will.

The concept of **free will** is a paradox that puzzled Samuel Johnson (1709-1784) [1] more than 200 years ago, as it had puzzled Lucretius[3] (100-50 BCE) [2] more than 2000 years ago. Johnson expressed the free will paradox succinctly: "All theory is against the freedom of the will; all experience is for it."

Stanislas Dehaene [3] resolved the paradox with the following informal[4] statement:

> "Our states are clearly not uncaused and do not escape the laws of physics – nothing does. But our decisions are genuinely free whenever they are based on a conscious deliberation that proceeds autonomously, without any impediment, carefully weighing the pros and cons before committing to a course of action. When this occurs, we are correct in speaking of a voluntary decision – even if it is, of course, ultimately caused by our genes [and circumstances]."

We arrived at our more formal solution to the paradox independently of Dehaene – and in agreement with him. Our resolution is based on the perspective of **Theoretical Computer Science** (**TCS**), particularly the fact that *computation takes time*. The fact that computation takes time was always known, of course, but computation time was never viewed as crucial for solving deep philosophical problems like those of consciousness and free will.[5]

Our more formal solution applies to a robot with a CTM brain[6] living in a deterministic world. For example, Conway's Game of Life[7] is such a world, and can support such a robot.

---

[1] mblum@cs.cmu.edu

[2] lblum@cs.cmu.edu

[3] Lucretius in *De Rerum Natura*: "If all movement is always interconnected, the new arising from the old in a determinate order – if the atoms never swerve so as to originate some new movement that will snap the bonds of fate, the everlasting sequence of cause and effect - what is the source of the free will possessed by living things throughout the earth?"

[4] Dehaene's statement is informal. For one, not all terms are defined: What is a "state"? What is a "conscious deliberation? What is the meaning of "a conscious deliberation without any impediment"?

[5] In his essay, "Why Philosophers Should Care About Computational Complexity" [12], Scott Aaronson gives many examples that address the significance of resource limitations, but never mentions "free will". While he mentions "consciousness" a few times, he does not pursue it.

[6] Our CTM is a probabilistic device. In a deterministic world, however, we want CTM to be deterministic. To solve the problem of converting a probabilistic CTM into a deterministic one, we introduce a pseudorandom generator (including a random seed) into the description of CTM. Given that CTM has a finite lifetime, T, this pseudorandom generator can produce all the "randomness" that CTM will ever require [14].

[7] See the 2022 Wikipedia article on Conway's Game of Life, which refers to [11]).

May 2024

In [4] we define a **Conscious Turing Machine** (**CTM**) that goes through life making decisions, some consciously, others unconsciously. During whatever *time* CTM consciously evaluates several options, before committing to any one of them, it can rightly claim to have free will. That's because whenever CTM does computations to arrive at a decision, it *"knows"* (has *conscious knowledge*) that it has at least two options to choose from.

Specifically, at each and every **clock tick**[8], the CTM's **Long Term Memory** (**LTM**) processors each submit a **chunk**[9] of information - an idea, a query, an answer to a query, a comment - to a formal **competition** for admission to CTM's **Short Term Memory** (**STM**). The latter is a tiny buffer (the stage) that at each clock tick receives the current **winning chunk (a conscious thought)** and immediately **broadcasts** it (sends the chunk to all LTM processors, the audience of experts). Those processors receive the chunk at the clock tick immediately following its broadcast, at which time the CTM is said to pay **conscious attention** to that chunk.

In what follows, except where we say otherwise, we assume that the world is deterministic. In a deterministic world, there is clearly no free will. What is surprising is the solution to the free will problem - that a (deterministic) robot in a deterministic world can rightly firmly believe in its own free will.

Our TCS perspective addresses **the paradox of free will** with the following definitions of **free will, the feeling of free will,** and the **exercise of free will** *in* the **CTM:**

- **Free will** is the ability to violate physics.

  As violating physics is impossible,[10] no animal, machine, or **CTM** has free will.

- **The feeling of free will** is the **conscious thinking**[11] – the sequence of *broadcasted conscious thoughts* – that occurs during the time that the **CTM**, faced with a choice, computes the consequences and utility of its possible actions in order to choose the action that suits its goals best.

- **The exercise of free will** is the act of choosing between two or more options, whichever option is believed to have the greatest utility. This is something that any animal or machine with a **CTM** brain can do.

  In this paragraph, the game of chess is a stand-in for any game that requires CTM to make a decision in a time greater than one tick of the clock - and such games exist [5]. Now consider a **CTM** that is called upon to play a given position in the game. Different processors suggest different moves. A **CTM** game playing processor indicates, by broadcast from **STM**, that it recognizes it has a choice of possible moves and that the decision which move to make merits a careful look at the consequences of each move. At the time the **CTM** recognizes the decision it must make, until the time it settles on its final decision, the **CTM** can *consciously* ask itself by broadcast from **STM**, "Which move should I make, this or that?", "If I do this, then what?", and so on. Thus, by *conscious thinking*, **CTM** converges on whichever move it reckons best. This has been defined above as the *exercise of free will*.

---

[8] **Time** in CTM is measured in **clock ticks, t = 0, 1, 2**, ..., **T**.  T is the (intentionally) finite lifetime of a CTM.

[9] A **chunk**, defined formally in [4], contains a "small" amount of information, a so-called the gist.

[10] Violating the laws of physics in a deterministic world (like Conway's Game of Life [11]) is clearly impossible. It is also impossible in the quantum world, where all actions are random selections from precise deterministically-defined (physicists call them calculable) probability distributions.

[11] In the **CTM**, *conscious thinking* is computation that involves entry to and broadcast from **STM**. Unconscious thinking is computation that does not involve **STM**: it is done entirely within LTM processors and through links. An example of *conscious* versus *unconscious thinking*: Learning to ride a bike requires conscious thinking. Once learned, this thinking becomes largely unconscious though **links** between **LTM** processors.

But will the **CTM "feel"** – in the way "feeling" is normally understood – that it has **free will**?

1. Consider the moment that the **CTM** asks itself "What move should I make?" meaning this query has risen to **CTM's STM** and, through broadcast, has reached the audience of **LTM** processors. In response, some **LTM** processors generate suggestions for what to do next, perhaps what move to pursue, each submitting its suggestion to an ongoing competition for **STM**. The suggestion that reaches **STM** gets broadcast.

2. The sequence of *thoughts* (comments, questions, suggestions and answers) that are broadcast from **STM** globally to all **LTM** processors gives the **CTM** *conscious knowledge* **of its control**: If the **CTM** is asked how it generated a specific suggestion, i.e., what thinking went into making that suggestion, its **Speech** or **Inner Speech** processor [4] would be able to articulate the fraction of conversation that was broadcast from **STM** (though perhaps not much more than that in the short term).

3. Many **LTM** processors compete to produce the **CTM**'s final decision, but **CTM** *consciously knows* only what got broadcast from **STM**, which is not all that was submitted to the competition. Moreover, most of **CTM**'s processors are not privy to the unconscious chatter (through **links**) among the (other) processors. To the **CTM**, enough is *consciously unknown* about the process that the decision can appear at times to be *plucked from thin air*. Even so, although **CTM** does not *consciously know how* its decisions were arrived at, except for what is in the high level broad strokes broadcast by **STM**, the **CTM** *can rightly take credit for making its decisions* (after all, it can tell itself they did come from inside itself), can explain some of those decisions with high level stories, and as for what it cannot explain, it can say "I don't know" or "I don't remember."

It is the *conscious knowledge* that:
a. there are choices,
b. the **CTM** can evaluate those choices (to the extent they become *consciously known* and that time allows), and
c. the **CTM** can rightly take credit for its decisions,

that generates the "*feeling*" of free will.

As for how **time** enters into the picture:

> In a deterministic world, the path to be taken is determined.
> In a quantum world, the probability distribution of paths is determined. So...
> > **"All theory is *against* the freedom of the will."**

> But when a problem is presented to the **CTM**, it does not *know* what path it will take.
> To *know*, it must compute, and computation takes time. And during that time, the **CTM** *knows* it will choose whichever option it decides is best.
> > **"All experience is *for* the freedom of the will."**

We now elaborate on Dehaene's explanation of free will, especially on his suggestion that free will is based on "a conscious deliberation that proceeds autonomously". What does it mean for a deliberation to be conscious? Does a deliberation need to be conscious in order for a feeling of free will to be generated? (Yes!) What makes a deliberation autonomous?

In the **CTM, conscious deliberation** is *conscious thinking*, a sequence of *broadcasted conscious thoughts*. Consequently, all processors, including those responsible for generating the "feeling" of free will, will *know* about the deliberation.[12] What it means for a deliberation to be **autonomous** is that the **CTM** makes its decision based on what it *knows*. Suppose it *knows* that it must do something at risk of some terrible consequence such as "pain" or "death". Is the decision to do the thing autonomous? Yes.[13]

---

[12] Processors responsible for the "feeling" of free will include the **Model-of-the-World** processor, the **Inner Speech processor**, etc.

Contemporary researchers including mechanical engineer/physicist Seth Lloyd [6], physicists Max Tegmark [7] and Sean Carroll [8], philosopher/neuroscientist Sam Harris [9], and computer scientist Judea Pearl [10], take a stance on free will kindred in spirit to ours.[14]

Our arguments for the "feeling" of free will work as well for a **CTM** in a deterministic world (like [11]) as for a **CTM** in a probabilistic or quantum world. In particular, *quantum weirdness* is totally unnecessary for explaining the feeling of free will.

### Acknowledgements

We gratefully acknowledge Michael Xuan and UniDT for their long-standing support of our work.

---

[13] A deliberation is **not autonomous** when an external operator controls **CTM**'s decisions so that **CTM** *cannot* do otherwise. This can be engineered by giving **CTM** a *non-standard* processor whose chunks can have *infinite* weight. Like Robby the Robot, in **Forbidden Planet** [13] that chunk can give **CTM** a "seizure" whenever it (the **CTM**) tries to do something forbidden.

[14] That said, we do not agree with Lloyd that "the intrinsic computational unpredictability of the decision-making process is what gives rise to our impression that we possess free will." While it is true that a final decision is not *consciously known* until it is made, we claim it is not the unpredictability that generates the "feeling" of free will. Rather, it is the recognition that there are choices, and that continuing deliberations (computations) are needed to select one.